\documentclass[twocolumn,floats,aps,superscriptaddress,nofootinbib]{revtex4}
\usepackage{times} 
\usepackage{amssymb,amsmath}
\usepackage{csquotes}
\usepackage[usenames,dvipsnames]{color}
\usepackage{epstopdf}
\usepackage[pdftex]{graphicx}
\usepackage[pdftex]{hyperref}
\hypersetup{a4paper,
    pdftitle={What is active wetting?} 
    pdfauthor={Uwe Thiele},  
pdfproducer={lateX},
pdfview=FitV,       
pdfstartview=FitB,
linkcolor=blue,     
citecolor=blue,     
urlcolor=red,      
breaklinks=true,    
colorlinks=true,
citebordercolor=0 0 0,  
filebordercolor=0 0 0,
linkbordercolor=0 0 0,
menubordercolor=0 0 0,
urlbordercolor=0 0 0,
pdfhighlight=/I,
pdfborder=0 0 0,   
bookmarksopen=true,
bookmarksnumbered=true
}

\DeclareGraphicsExtensions{.jpg, .pdf, .tif, .png}
\graphicspath{{.},{./Figures/}}  
\usepackage[usenames,dvipsnames]{color}
\usepackage[normalem]{ulem}

\newcommand{\bfuwe}[1]{\textbf{\textcolor{BrickRed}{#1}}}
\begin{document}
%
\title{What is active wetting?}
\author{Uwe Thiele}
\email{u.thiele@uni-muenster.de}
\homepage{http://www.uwethiele.de}
\thanks{ORCID ID: 0000-0001-7989-9271}
\affiliation{Institute of Theoretical Physics, University of M\"unster, Wilhelm-Klemm-Str.\ 9, 48149 M\"unster, Germany}
\affiliation{Center for Data Science and Complexity (CDSC), University of M\"unster, Corrensstr.\ 2, 48149 M\"unster, Germany}
\begin{abstract}
  In recent years the term \textit{active wetting} has gained some traction in works describing, analyzing and modeling a wide variety of wetting phenomena, for instance, in the contexts of biomolecular condensates, of cell layers or cell aggregates, and of active Brownian particles. The present perspective discusses a coarse classification of wetting phenomena that accounts for this. First, different categories of static and dynamic wetting of passive liquids are briefly introduced, in particular, distinguishing equilibrium wetting, relaxational wetting, driven wetting, and reactive wetting. Second, an overview is given of the various phenomena recently described as active wetting. We conclude by discussing a possible definition of active wetting together with a number of caveats that one might want to keep in mind when using such classifications.\\[2ex]
\bfuwe{Published as \textit{The European Physical Journal E} \textbf{49}, 56 (2026).  doi: \href{http://dx.doi.org/10.1140/epje/s10189-026-00590-y}{10.1140/epje/s10189-026-00590-y}}  
 \end{abstract}
\maketitle
%

\section{Introduction} \label{sec:intro}
%
The interest in static and dynamic wetting phenomena has a rather long history \cite{Hard1922n,PoVi2006pt}, see, e.g., \cite{Hauk1713ptc,Segn1751csrsg,Youn1805ptrs,Laplace1806} and query~31 in Ref.~\cite{Newton1730}. Wetting, in its simplest form, refers to the interaction between a liquid, an ambient fluid phase (often gas), and a rigid smooth homogeneous solid substrate. On a more abstract level wetting phenomena occur in any situation where more than two phases meet. The present understanding of these phenomena for classical molecular liquids is systematically presented in many books and reviews, e.g., Refs.~\cite{RowlinsonWidom1982,Genn1985rmp,LeJo1992rpp,GennesBrochard-WyartQuere2004,StarovVelardeRadke2007,BEIM2009rmp,CrMa2009rmp,Brutin2015,Bormashenko2017}. 

However, an increasing number of studies addresses phenomena that they refer to as \textit{active wetting}.  Although definitions are often left implicit, a multi-faceted concept of active wetting is emerging that involves, e.g., active soft matter -- characterized by self-propelling constituents and/or active stresses \cite{MJRL2013rmp,Nage2017rmp,BFMR2022prx,VrWi2025epje}. Because the notion of active wetting is fast spreading although it lacks a clear definition, the present contribution aims at providing a tentative classification of wetting phenomena that incorporates systems featuring active wetting. As a first step, we introduce a general distinction of four categories of wetting phenomena, namely, equilibrium wetting, relaxational wetting, driven wetting, and reactive wetting -- all studied with classical molecular liquids, i.e., with passive not active liquids.\footnote{One commonly defines \textit{active matter} as \enquote{composed of self-driven units, active particles, each capable
of converting stored or ambient free energy into systematic movement} \cite{MJRL2013rmp}, also see \cite{EESS2000epjb,RaSi2006ssc}. For more details about the notion see the introduction of \cite{BFMR2022prx}, section~6 of \cite{Pismen2023} and section~9 of \cite{SaarloosVitelliZeravcic2024}. For a recent critical reflection see \cite{VrLC2025preprint}.}
A few other well established terms like forced wetting and adaptive wetting are also discussed in the context of this scheme. \footnote{Note that most reviews related to wetting phenomena only distinguish static and dynamic wetting, and do not intend to give a more differentiated general classification. Ref.~\cite{Genn1985rmp} discusses statics and dynamics of wetting (without formal definition), a distinction that is also followed by \cite{BEIM2009rmp,AnSn2020arfm} where also forced or driven contact lines are specifically considered, \cite{TeDS1988rpap} distinguishes the spontaneous and forced motion of dynamic contact lines, \cite{CrMa2009rmp} mainly discusses dynamic wetting phenomena in the context of dewetting. Note that static contact angles that differ from equilibrium values in certain out of equilibrium systems seem not to fit into any of these schemes.}

Second, we briefly review a selection of recent works that use the notion of active wetting to describe certain observed phenomena in a variety of experimental and theoretical contexts, e.g., including inner structures of biological cells, cell aggregates and monolayers, and clusters resulting from motility-induced phase separation of active Brownian particles.

Finally, we offer a tentative definition of active wetting that builds on the previously discussed notions of equilibrium, relaxational, driven, and reactive wetting. We also explain that there exists a number of caveats that speak against a rigid classification into a small  number of categories. Any classification of natural phenomena will depend on the employed conceptual idealizations and resulting mathematical approximations in their theoretical treatment. Our conclusion is that one might indeed distinguish phenomena of equilibrium, relaxational, driven, reactive, and active wetting but should always carefully clarify which characteristics of the considered system one refers to and which idealizations are used.

\section{Equilibrium wetting}
\label{sec:equi}

\begin{figure}[tbh]
\includegraphics[width=1.0\hsize]{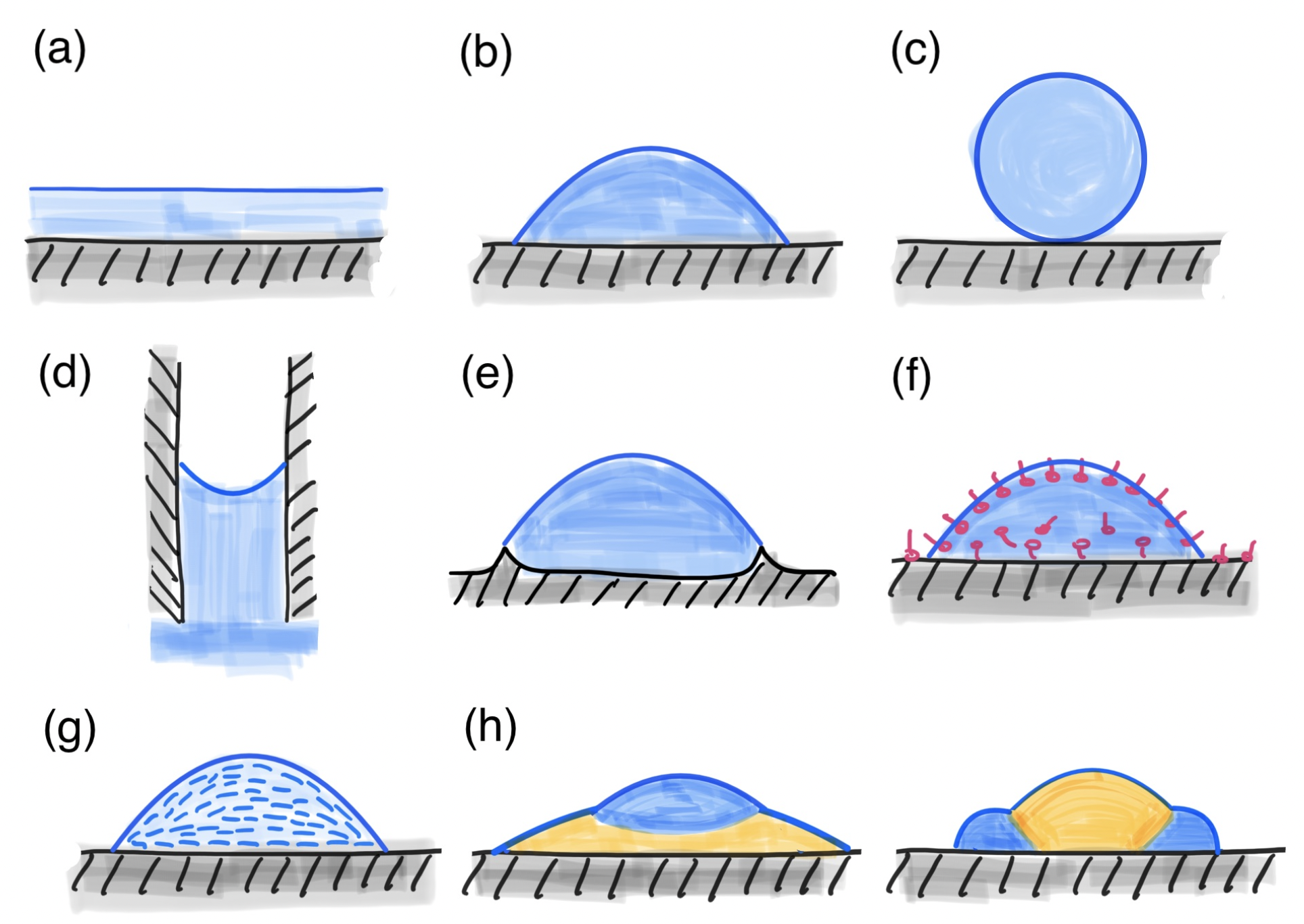}
\caption{Sketches of equilibrium wetting: (a) complete wetting, (b) partial wetting, and (c) non-wetting for a liquid on a rigid smooth solid substrate; (d) equilibrated capillary rise, (e) drop of partially wetting liquid deforming a soft solid substrate, (f) surfactant-laden drop in an equilibrated state with steady respectively uniform but in general different surfactant concentrations in liquid bulk and at the three interfaces, (g) drop of nematic liquid crystal with parallel anchoring at all interfaces, and (h) two different equilibrium states for a drop of a phase-separating liquid mixture.
}
\label{fig:sketch-equilibrium}
\end{figure}

The \textit{equilibrium wetting} scenario applies when the system is in a thermodynamic equilibrium state, defined as minimum of an appropriate thermodynamic potential, e.g., a free energy or a grand potential. Macroscopically, the type of equilibrium is determined by the energies of the involved sharp interfaces, that themselves depend on temperature, pressure, etc. 
One distinguishes (i) complete wetting (extended uniform liquid film between substrate and ambient phase, spreading coefficient $S=\gamma_\mathrm{sg}-\gamma_\mathrm{sl}-\gamma_\mathrm{lg}\ge0$, equilibrium three-phase contact angle $\theta_{\mathrm{eq}}=0$, see Fig.~\ref{fig:sketch-equilibrium}~(a)), (ii) partial wetting (sessile liquid drop of spherical cap shape on the uniform solid rigid substrate, $-2\gamma_\mathrm{lg}<S<0$,  $0<\theta_{\mathrm{eq}}<\pi$, see Fig.~\ref{fig:sketch-equilibrium}~(b)), and (iii) non-wetting (ideal liquid sphere on substrate, $S\le-2\gamma_\mathrm{lg}$,  $\theta_{\mathrm{eq}}=\pi$, see Fig.~\ref{fig:sketch-equilibrium}~(c)). Here, $\gamma_\mathrm{lg}$, $\gamma_\mathrm{sl}$, and $\gamma_\mathrm{sg}$ are the interface energies per area of the liquid-gas, solid-liquid, and solid-gas interface, respectively.  Spreading coefficient, equilibrium contact angle and interface energies are related by Young's law $\gamma_\mathrm{lg}\cos \theta_{\mathrm{eq}} = \gamma_\mathrm{sg}-\gamma_\mathrm{sl}$ and the equivalent formulation as Young–Dupr\'e law $S=\gamma_\mathrm{lg}(\cos \theta_{\mathrm{eq}}-1)$. For details see, e.g., section~II of \cite{Genn1985rmp} or section~I.B of \cite{BEIM2009rmp}.

The shift between partial and complete wetting that can occur when temperature or other control parameters are changed corresponds to a wetting transition -- a phase transition that may be of first or second order (section~III of \cite{Genn1985rmp}, section~II of \cite{BEIM2009rmp}). One may also say a wetting transition is related to the unpinning (or unbinding) of a liquid-fluid interface from the substrate \cite{Hinr2006pa}. The change may involve further pre-wetting and layering transitions, where states appear that consist of thin layers of liquid of defined thicknesses that cover the substrate \cite{Diet1988,Inde2010pa}, also see the unbinding transitions discussed in Ref.~\cite{PRJA2012prl}. Notably, topographical and chemical substrate roughness and texture may influence the wetting behavior even causing super-hydrophobicity \cite{LeLi1998prl,BaDi2000pre,BiTQ2002csa,LaQu2003nm,TBBB2003epje,Quer2008armr,SaPK2011jfm,WWSN2019l}.

If other energetic influences are present or if the liquid is complex, the equilibrium state may depend on more aspects. Examples include the equilibrium height  of a meniscus in a vertical capillary if gravitation is present (Fig.~\ref{fig:sketch-equilibrium}~(d)) \cite{GennesBrochard-WyartQuere2004}, the deformation of a soft solid substrate (Fig.~\ref{fig:sketch-equilibrium}~(e)) \cite{AnSn2020arfm}, and the contact angle-changing equilibrium adsorption of surfactants at all interfaces for a surfactant-laden drop (Fig.~\ref{fig:sketch-equilibrium}~(f)) \cite{TSTJ2018l}. For complex liquids like solutions, suspensions, mixtures, liquid crystals etc.\ their inner degrees of freedom may influence the interfacial energies and therefore directly the wetting behavior. It may be further changed due to related additional bulk energy contributions that allow for more intricate phase behavior.  Examples include the the interplay of elasticity, interface anchoring and director orientation-dependent interface energy for liquid crystals in the nematic (or other) phases \cite{Rey2000lc,Rey2003pre,Rey2007sm,CDRS2011acis}, see Fig.~\ref{fig:sketch-equilibrium}~(g), the interplay of interfaces with the bulk decomposition of a mixture \cite{WoSc2004jpm,Onuk2009jcp} itself related to equilibrium configurations of compound drops \cite{MaAP2002jfm,BSNB2014l,ZGLD2021jfm,Kita2024jns,DiTh2025prf} (Fig.~\ref{fig:sketch-equilibrium}~(h)), and the leak-out transition observed for polymer solutions \cite{Boud1987jp,FoBr1997el,BrFB2000ijes,NuST2023csaea}. This implies also potentially more intricate behavior in the dynamic cases discussed next.
 
 \section{Relaxational wetting}
 \label{sec:relax}
 \begin{figure}[tbh]
\includegraphics[width=1.0\hsize]{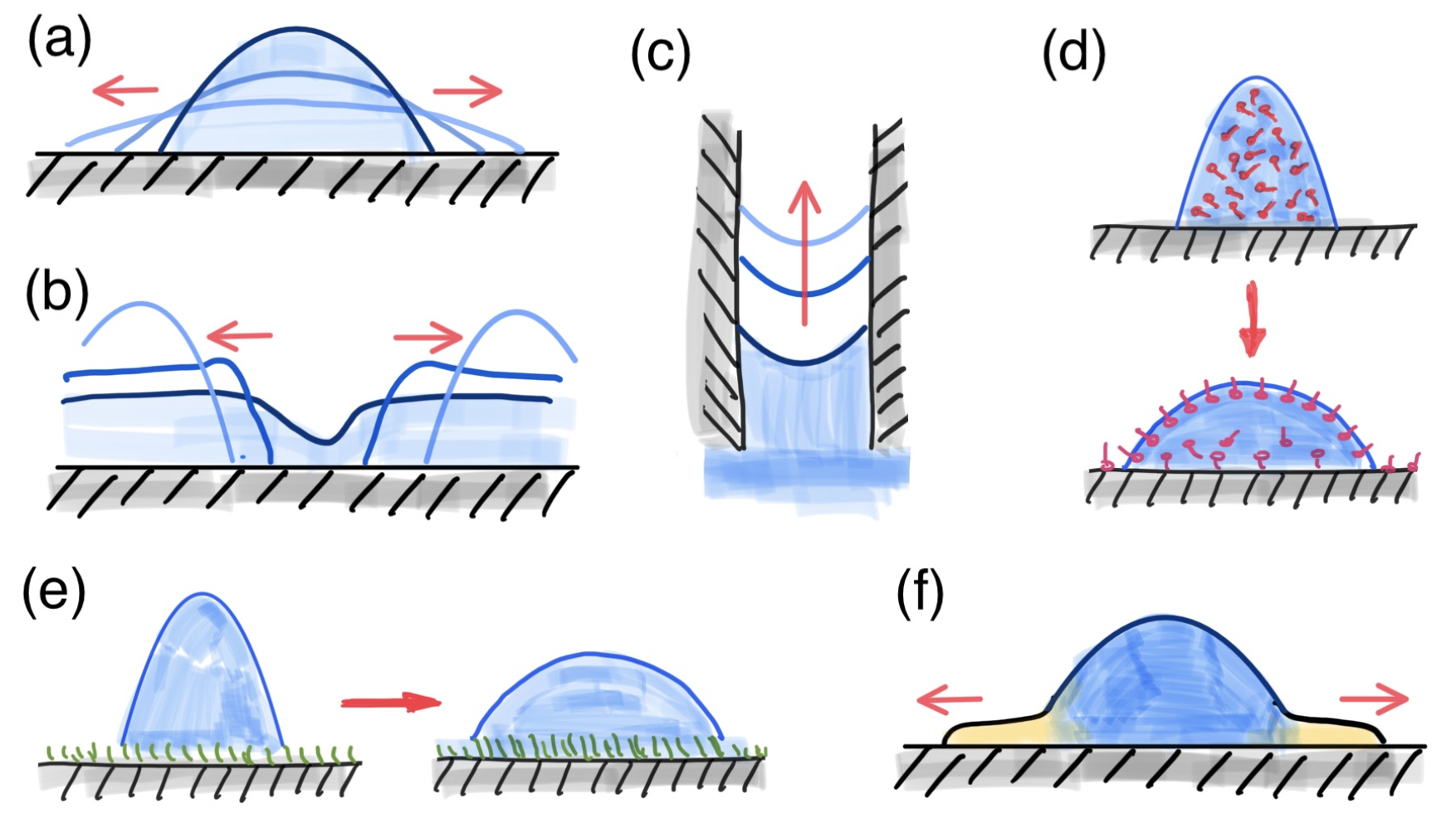}
\caption{Sketches of relaxational wetting: (a) spreading drop, (b) dewetting liquid film via nucleation of a hole, (c) the process of capillary rise, (d) coupled drop spreading and surfactant adsorption dynamics, (e) spreading and imbibition dynamics for a drop on a polymer brush, and (f) spreading of a single-component adsorption layer from a drop of mixture (leak-out).
}
\label{fig:sketch-relaxational}
\end{figure}

In addition to equilibrium wetting, it is crucial to consider the dynamic aspects of wetting both in situations where an out-of-equilibrium state relaxes toward an equilibrium state and in situations where an external driving force keeps the system permanently out of equilibrium. In the present section we discuss the former case that we term \textit{relaxational wetting} while the latter case is considered in section~\ref{sec:driven}.  In relaxational wetting a sessile drop or another configuration that may or may not involve a three-phase contact line is initiated in an out-of-equilibrium state, e.g., the initial contact angle differs from the equilibrium contact angle $\theta_{\mathrm{eq}}$. Without externally imposed additional force the system approaches an equilibrium state (a minimum of the appropriate energy functional, e.g., of the free energy in an isothermal setting without particle exchange with the environment) while dissipating the surplus free energy (section~IV of \cite{Genn1985rmp}, section~III of \cite{BEIM2009rmp}).

For instance, a liquid drop of nonvolatile liquid placed on a smooth solid substrate spreads (or retracts) till reaching $\theta_\mathrm{eq}$ (Fig.~\ref{fig:sketch-relaxational}~(a)). If the liquid completely wets the substrate, $\theta_{\mathrm{eq}}=0$, the relaxation process follows a power law: Tanner's law $R \sim t^{1/10}$ where $R$ is the drop radius and $t$ time \cite{Tann1979jpd} and the equilibrium state - an infinitely extended, infinitely thin liquid film is only reached after an infinite amount of time. If $\theta_\mathrm{eq}$ is finite the equilibrium state is approached exponentially slowly. If the substrate is rough or patterned, additional complexities arise due to the many metastable states and the related contact angle hysteresis \cite{ScWo1998prl,CuFe2001el,VeSK2011pre,BLKS2022cocis}.

Also the dewetting of an initially uniform liquid film (that is rendered unstable by a temperature quench) either via spinodal dewetting or nucleation (Fig.~\ref{fig:sketch-relaxational}~(b)) represents an example of relaxational wetting dynamics where initial film rupture, growth of individual holes, transversal instability of dewetting fronts, and the coarsening of ensembles of small drops into fewer larger drops are studied in detail (see e.g.\ section~V of \cite{CrMa2009rmp} and \cite{Thie2003epje,PHMJ2019pnasusa,KGDF2020arfm,KoSK2022pt}). Another example is capillary rise in thin tubes \cite{ScWo2000pre} (Fig.~\ref{fig:sketch-relaxational}~(c)).

More recently, considerations of equilibrium wetting and relaxational wetting have been extended toward deformable elastic and adaptive substrates \cite{BiRR2018arfm,BBSV2018l,CBGS2018cocis,AnSn2020arfm,MeBS2021sm,EDSD2021m,HDGT2024l,HNBP2024sm}. One may use the term \textit{adaptive wetting} for all such cases as the substrate adapts its topography by deformation and its wettability by absorption. Other types of substrate restructuring may combine these effects \cite{BBSV2018l}. The category includes porous and poro-elastic substrates that change their wetting properties under liquid imbibition \cite{Star2004acis,Gamb2014cocis,JoTS2019ci,HaTh2025prf,FPEB2023prl}. When a drop spreads on a polymer brush (Fig.~\ref{fig:sketch-relaxational}~(e)), liquid imbibes the polymer brush what changes its wettability and thickness profile \cite{HDGT2024l}.
If complex liquids are involved the relaxation of the drop profile and the contact line is coupled to the relaxation of other degrees of freedom, i.e., there exist additional channels of dissipation. For instance, Fig.~\ref{fig:sketch-relaxational}~(d) sketches a surfactant-laden drop that spreads while surfactant adsorbs at all interfaces, see e.g.\ \cite{Star2004acis}. When a mixture spreads the bulk drop may spread slower than a nanoscopic adsorption layer of one component that is leaking out (Fig.~\ref{fig:sketch-relaxational}~(f)).  The additional degrees of freedom may also  give rise to novel instabilities, e.g., spreading drops of liquid crystals \cite{PoCa2005l} and of surfactant solutions \cite{CaCa1999l,HCPC2004csaea,MaCr2009sm} show front instabilities not observed for simple liquids.

\section{Driven wetting}
\label{sec:driven}
\begin{figure}[tbh]
\includegraphics[width=0.9\hsize]{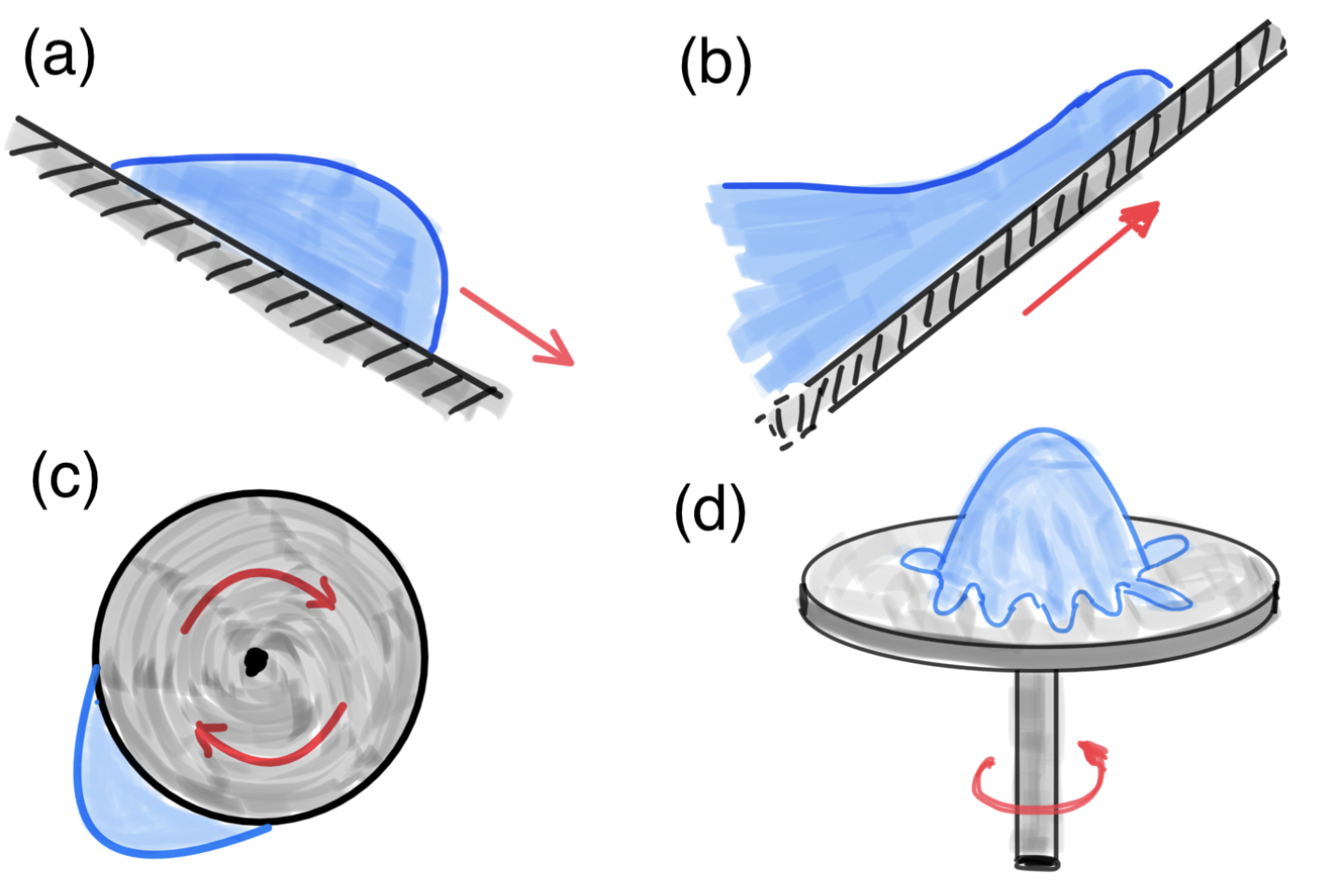}
\caption{Sketches of driven wetting: (a) drop sliding down an incline, (b) foot (or film) profile drawn out of a liquid bath by a moving plate (Landau-Levich geometry), (c) pending drop on the outside of a rotating cylinder, and (d) drop on spinning plate showing a contact line instability.
}
\label{fig:sketch-driven}
\end{figure}

In contrast to relaxational wetting, in \textit{driven wetting} (or \textit{forced wetting}) a global breaking of spatial parity (i.e., reflection symmetry w.r.t.\ a plane orthogonal to the substrate) by external lateral forces or fluxes keeps an entire drop or a single contact line permanently out of equilibrium (see chap.~IV of \cite{Genn1985rmp}, \cite{TeDS1988rpap,SnAn2013arfm,HuSc1971jcis}).  Mechanical and thermodynamic forces may result from imposed global gradients of pressure, temperature, wettability, potential energy, etc. For instance, drops slide down an ideally smooth incline driven by the downhill force resulting from the gradient in potential energy \cite{PoFL2001prl,EWGT2016prf,BBDT2025nrp} (Fig.~\ref{fig:sketch-driven}~(a)), or are deformed or dragged along when pending on the outside of a rotating cylinder \cite{Pukh2005smj,GrKe2009pf,LRTT2016pf} (Fig.~\ref{fig:sketch-driven}~(c)).
Contact lines advance and become transversally unstable due to centrifugal forces for a sessile drop on a spinning substrate \cite{MeJF1989prl} (Fig.~\ref{fig:sketch-driven}~(d)), due to gravity on an incline \cite{DiKo2001prl}, or due to Marangoni forces at the liquid-gas interface caused by an imposed temperature gradient along the substrate \cite{CHTC1990n}. Other examples are drops that move due to an imposed gradient in wettability, softness, or humidity \cite{ChWh1992s,BBJG2017eml,ZhQi2022prf,BROW2024epje}. Besides imposing the driving force one may also impose the (mean) velocity of the contact line along the substrate, e.g., in the Landau-Levich geometry \cite{RiBo2017aci,TWGT2019prf} (Fig.~\ref{fig:sketch-driven}~(b)), for the rotating cylinder (Fig.~\ref{fig:sketch-driven}~(c)), and in Langmuir-Blodgett transfer \cite{Blod1935jacs,SpCR1994el,KGFT2012njp,OlCA2022cr}, or by inflating/deflating a droplet via a syringe pump \cite{LWLH2002acis,PuSe2008l,SHNF2021acis}. %
\footnote{For the radially symmetric set-ups the \enquote{parity breaking} refers to parity along the radial direction, e.g., a centrifugal force for the drop on a spinning substrate resulting in driven spreading or a central in- or outlet forcing advancement or retraction of the circular contact line, respectively.} In the Landau-Levich case a plate is moved out of a liquid bath thereby dragging liquid into a prolonged foot-like meniscus or wetting film covering the plate. The configuration or film thickness depends on the speed of withdrawal --  a phenomenon central to many coating processes \cite{Rusc1985arfm,RiBo2017aci}. 

In such ways one can keep the system permanently out of equilibrium, thereby allowing for phenomena that can not occur during a relaxational process. This includes time-periodic states like self-sustained oscillations related to stick-slip motion of the contact line on flexible or adaptive substrates \cite{PuSe2008l,KDNR2013sm,PBDJ2017sm,GASK2018prl,MoAK2022el,GrHT2023sm}, involved scenarios of depinning for drops on rotating cylinders \cite{Thie2011jfm,LRTT2016pf}, and drop merging-splitting cycles when a periodic array of drops slides down an incline \cite{PoFL2001prl,LeDL2005jfm} where even the classical period-doubling route to chaos can be encountered \cite{EWGT2016prf} as also known from a dripping faucet \cite{DrHi1991ajp}. In the Landau-Levich setting, dynamic wetting and unbinding transitions occur \cite{ZiSE2009epjt,GTLT2014prl} that are externally driven equivalents of corresponding equilibrium transitions \cite{PRJA2012prl}. Note that driven drop and contact line motion on irregularly or regularly structured substrates is also frequently studied, see e.g.\ \cite{JoRo1990jcp,QuAD1998l,ThKn2006njp,Quer2008armr,BKHT2011pre,SaPK2011jfmb,HTHB2012el,SaKa2013jfm,VFFP2013prl,SeBr2014sm,ShTO2025l}. Also, sliding on adaptive substrates is considered \cite{HDGT2024l,BHVS2024l,ZWLS2024am}, and forcibly sliding drops are now even employed as probes in a technique to determine spatially resolved maps of the lateral adhesion force (scanning drop friction force microscopy) \cite{HLSW2022l}.

In all cases discussed so far in sections~\ref{sec:relax} and~\ref{sec:driven}, contact lines move and therefore have modified properties as compared to the equilibrium case in section~\ref{sec:equi}. For instance, the contact angle is not given by Young's law anymore but is amended by the nonequilibrium conditions as captured by laws that relate the dynamic contact angle $\theta_\mathrm{dyn}$ with the velocity $U$ of the contact line \cite{Moha2022jap}, e.g., the hydrodynamic Cox-Voinov law $U\sim\theta_\mathrm{dyn}^3-\theta_\mathrm{eq}^3$ \cite{Voin1976fd,Hock1983qjmam,Cox1986jfm,SnAn2013arfm,LuGa2025jfm} or the law $U\sim \cos \theta_\mathrm{eq}-\cos \theta_\mathrm{dyn}$ derived based on a variational energy-dissipation principle \cite{Pesc2018pf} and also emerging from the molecular kinetic model close to equilibrium \cite{Blak2006jcis,BFDD2015pf}.

Beside the described clear-cut cases of driven wetting related to parity breaking by imposed global lateral forces one could also include another type of driven (de)wetting, namely, where parity is broken by an initial condition that corresponds to a front between different steady states, e.g., to a front between two thermodynamic phases or between an unstable and a stable state \cite{Saar2003pr}. However, depending on the fine details of the employed definition such processes might also be classified as reactive wetting, see below.
For instance, imposing a constant uniform ambient chemical potential (partial vapor pressure, humidity) that promotes liquid evaporation one can force a contact line to continuously recede \cite{LyGP2002pre}. If in such a setting only the solvent of a solution or suspension evaporates, a continuously moving front of evaporative dewetting can result in a plethora of large-scale deposition patterns of solute \cite{Deeg2000pre,HaLi2012acie,Lars2014aj,Thie2014acis}, see Fig.~\ref{fig:sketch-reactive}~(b). Such driving by phase transitions can also occur when liquid drops move while consuming (or depositing) a solid layer of the same material \cite{LaRi2005prl,YoPi2005pre}. Actually, in a frame moving with the mean contact line speed such situations become very similar to the Landau-Levich case of a moving substrate, although the speed is self-selected and not externally imposed. 

\section{Reactive wetting} 
\label{sec:reactive}
\begin{figure}[tbh]
\includegraphics[width=0.9\hsize]{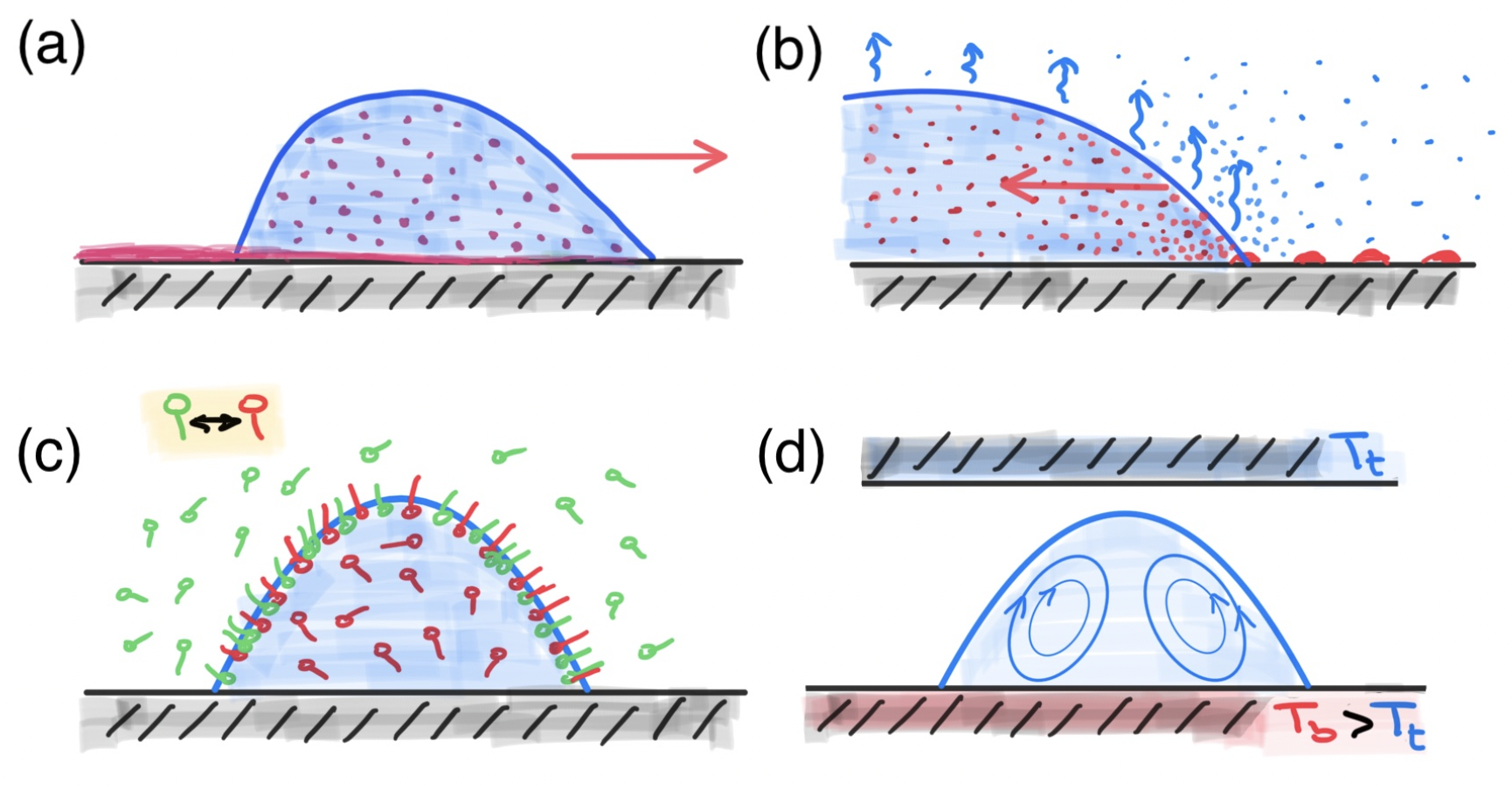}
\caption{Sketches of reactive wetting: (a) moving drop self-propelled by an adsorption reaction that renders the liquid-solid interface less wettable, (b) evaporative dewetting front for a drop or film of a suspension/solution leaving behind a patterned solute deposit, (c) drop with two species of chemically reacting surfactants coupled to two chemostats, and (d) drop on a heated substrate.
}
\label{fig:sketch-reactive}
\end{figure}

Alternatively, one may introduce a rather wide category of \textit{reactive wetting} that encompasses not only any (de)wetting process that significantly involves chemical reactions but also such processes that are caused or strongly influenced by other changes of state (in systems without externally imposed lateral forces or fluxes). With such an extension, reactive wetting would then cover evaporative dewetting fronts and the other cases mentioned in the final paragraph of section~\ref{sec:driven} but also alloying, physisorption, etc.

Common examples of reactive wetting involve, e.g., changes in substrate wettability that occur when a solute within a sessile drop chemisorbs at the substrate, when a liquid metal droplet alloys the topmost substrate layer, and when a self-assembled monolayer (SAM) is formed at a substrate underneath a drop of solution \cite{KuPr2007aci,CFBG2008l, YMSS2009jpm,ShBM2010e,EuVo2016jms,Tait2025f}. Also the dissolution of the substrate by the spreading liquid drop can be seen as an example where substrate material dissolves into the liquid due to a phase change or chemical reaction \cite{GoSh2004l,SSYM2012jms,EuVo2016jms}, this includes corrosion-induced droplet spreading \cite{RRJL2023sr}. One may then argue that the above-mentioned deposition of solutes in evaporative dewetting should also be included here -- a question we come back to in section~\ref{sec:conc}.

A particularly interesting example are chemically driven self-propelled drops that represent a self-organized dynamic state where the moving drop creates the wettability gradient needed to move at its speed \cite{DoOn1995prl,BrGe1995crassi,SKYN2005pre,JoBT2005epje,VoTh2024jem}, see Fig.~\ref{fig:sketch-reactive}~(a). Beside the reactions that change the surface energy of the substrate, i.e., its wettability, reactions may also involve surfactant molecules at the free liquid-gas interface thereby changing its surface tension and furthermore causing Marangoni forces due to created surface tension gradients \cite{ScSt1960n}. Examples include sessile drops with two species of reactive surfactants that can show a wide variety of dynamic wetting phenomena from simple spreading to involved emerging modes of oscillation and self-propulsion \cite{VoTh2024jem,VoTh2025prf,VoTh2026arxiv}, see Fig.~\ref{fig:sketch-reactive}~(c).

The similarity of dynamic wetting phenomena caused by chemical reactions and by phase transitions is illustrated by evaporation and other mass transfer processes that drive complex behavior like drop propulsion, pulsation, and rotation \cite{PiAn2014cocis} similar to phenomena described for the case of chemical reactions \cite{VoTh2026arxiv}. Here, one might then also include phase separation processes under the influence of chemical reactions if the latter influence the wetting behavior at confining walls, see e.g.\ the corresponding remarks in \cite{Zwic2022cocis}. We expect that, in general, the rich emerging physics of active emulsions \cite{WZJL2019rpp} results in intricate static and dynamic wetting behavior \cite{ZiZw2024jcp}. For instance, out-of-equilibrium chemical binding between droplet material and a solid substrate can significantly increase the parameter range where prewetting occurs as compared to an equilibrium system \cite{ZBHJ2021njp,ZLHJ2024njp,ZwPB2025rpp}. 

Tentatively, one could say that reactive wetting results from imposed persistent gradients that are not spatial along the substrate as in section~\ref{sec:driven} but between reservoirs of different properties. Examples include a temperature contrast between the substrate and the ambient phase (see Fig.~\ref{fig:sketch-reactive}~(d)), a chemical potential difference between drop bulk and vapor or drop bulk and drop surface, etc. Such a wide definition allows for all mentioned cases but then also includes the effect of superspreading that occurs for certain surfactant-laden drops because there the transition of surfactant molecules from the liquid-gas to the liquid-solid interface is a crucial ingredient \cite{KaCM2011jfm}. Further, reactive wetting will then also cover cases where energy or mass is transferred between the solid substrate and the ambient phase across the sessile drop or an liquid film. This then includes the intricate process of freezing liquid drops that may involve prewetting layers \cite{SLNA2021nc}, frost halos \cite{JuTP2012pnasusa}, and can result in pointy ice drops \cite{SnBr2012ajp}.

\section{Active wetting}
\label{sec:active}
\begin{figure}[tbh]
\includegraphics[width=1.0\hsize]{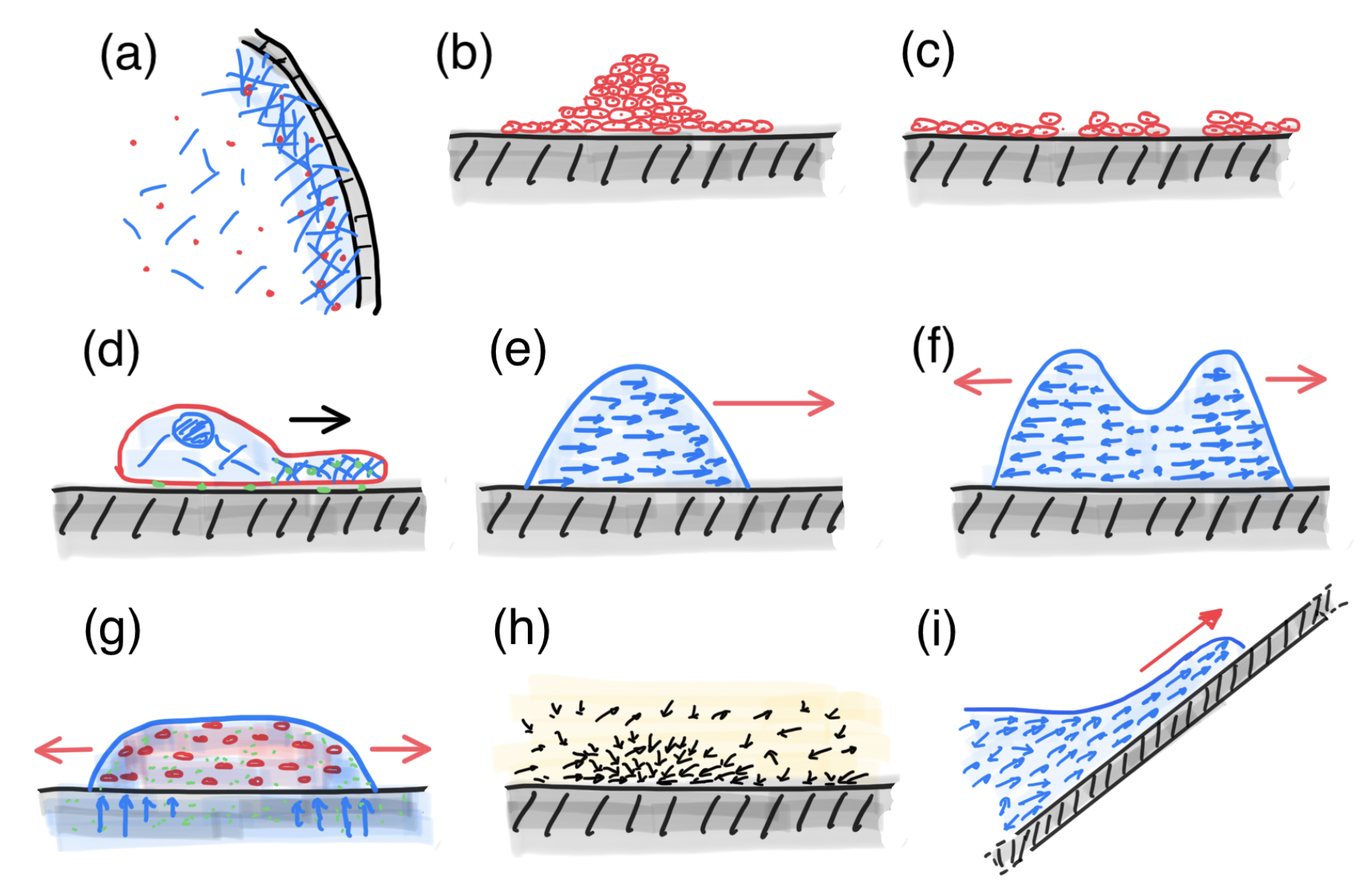}
\caption{Sketches of active wetting: (a) static active wetting layer at membrane, (b) resting cell aggregate and (c) cell monolayer on solid substrate, (d) individual crawling cell with protrusion, (e) moving and (f) splitting drop of active polar liquid, (g) osmotically growing biofilm, consisting of bacteria, extracellular matrix, nutrient-rich water, on solid agar, (h) partially wetting clusters of active Brownian particles at a solid wall, and (i) active liquid dynamically wetting (climbing up) a wall in a Landau-Levich geometry.
}
\label{fig:sketch-active}
\end{figure}

Having discussed equilibrium, relaxational, driven and reactive wetting we now turn to \textit{active wetting}. However, instead of starting with a definition, we first briefly review a selection of recent works that use the notion to describe certain observed phenomena. Afterwards we offer a tentative definition (again with some caveats). 

To our knowledge, the first work%
\footnote{Excluding mentions of \enquote{surface active wetting} in the context of SURFace ACTive AgeNTS (surfactants).}
that uses the term ``active wetting'' is Ref.~\cite{JKPR2013epje}. There, the actin cortex underlying the cell membrane of animal cells is discussed as an active wetting layer, i.e., a thin film of large uniform actin density that wets a substrate (the cell membrane) and whose thickness is determined by the balance of the rates of continuous actin fiber polymerization at the substrate and of fiber depolymerization in the film, see Fig.~\ref{fig:sketch-active}~(a). The contractile active stress, that results from corresponding chemical activity of molecular motors, in turn results in a nonmonotonic dependence of the effective osmotic pressure on actin concentration. The resulting effective thermodynamic instability yields a high-density wetting layer phase -- separated from the low-density bulk of the cell. The emergence of the layer at a critical strength of activity may be seen as a nonequilibrium analog of a (pre)wetting transition as mentioned in section~\ref{sec:equi}. Instabilities of a membrane substrate coupled to such an actin cortex are also considered, but without explicit mention of a wetting layer \cite{MSRR2014prl}. For the wider context of active gel theory, see \cite{MJRL2013rmp,BFMR2022prx}.

A second biophysical example is the adhesion and spreading behavior of drop-like cell aggregates \cite{DGND2011pnasusa,DoDB2012sm,BSKC2014pnasusa,PiAC2022epje} (Fig.~\ref{fig:sketch-active}~(b)) and of epithelial cell layers \cite{KoPi2013sm,KoPi2013pd,BEHR2013pnasusa,BlCa2017sm,BVBS2017sm,PABG2019np,MoYa2019np,TrBM2021sm,PaMM2026ebj} (Fig.~\ref{fig:sketch-active}~(c)) on soft and rigid solid substrates as well as on liquid substrates.\footnote{The surface tension of cell aggregates and their spreading was earlier studied, see e.g.\ \cite{FPFS1996d} and references therein.}
This is relevant, e.g., in embryonal development \cite{MGBK2017dc,WTYT2018bj} and for the onset of metastatic dissemination \cite{LMHB2025as}. In particular, Ref.~\cite{PABG2019np} investigates the wetting of epithelial tissues and discusses the transition between a three-dimensional sessile cell aggregate and an epithelial monolayer as an active equivalent of the thermodynamic wetting transition \enquote{whose physics differs fundamentally from that of passive wetting phenomena.} The difference lies in the presence of active stresses related to cell-cell and cell-substrate traction forces that living cells generate. The competing influences of the two types of stress result in an additional intrinsic length scale related to a critical monolayer radius for the wetting transition. Refs.~\cite{DGND2011pnasusa,DoDB2012sm} analyze the spreading of a cell monolayer from a sessile cell aggregate and find a wetting transition when increasing substrate stiffness or when changing the ratio of cell–substrate and cell–cell adhesion, also consider Ref.~\cite{AlCa2019l}. Fingering instabilities of advancing contact lines, i.e., of epithelial spreading fronts, are discussed in Refs.~\cite{KoPi2013sm,AlBC2019prl,TrBM2021sm,Aler2022jpat}, somewhat similar to the ones mentioned in section~\ref{sec:driven} for driven wetting. For a partial review on the statics and dynamics of active wetting, namely, in the context of biological cells and their aggregates see Ref.~\cite{PaMu2026arcmp}.

Active dewetting of cellular layers is also discussed in several biophysical contexts. For example, Ref.~\cite{DoBr2012epje} studies the dewetting of metastable cohesive cellular monolayers on nonadhesive substrates, Ref.~\cite{BlSh2023prl} discusses the aggregation of a bacterial biofilm into droplet-shaped mounds that develop into a fruiting body as a dewetting phenomenon, and Ref.~\cite{HHMS2024c} considers the dewetting of sub-epithelial (mesenchymal) cell layers that results in the formation of clusters that in turn promote villus folding in the intestine. Furthermore, layers of smooth muscle cells may transform into clusters in a process initiated by supercritical growing holes \cite{WGVS2023cb} (Fig.~\ref{fig:sketch-active}~(c)) similar to relaxational dewetting for layers of passive liquids.

Substrate softness may be employed to drive collective cell motion by durotaxis along stiffness gradients of soft substrates \cite{PPFG2023np}. Such gradient-driven wetting can also be caused by chemotaxis in gradients of extracellular signals \cite{FCWD2025pnasu}. Interestingly, the latter can even result in the shedding of daughter aggregates in a pearling instability of the receding contact line not unlike observations for (driven) sliding drops on an incline \cite{PoFL2001prl,EWGT2016prf} and for receding dewetting fronts \cite{ReSh2001prl}. Notably, active motion may also be guided by gradients in transport coefficients, see e.g., frictiotaxis and viscotaxis of cell migration \cite{GHWA2026sm}.
However, cell aggregates may even migrate without external gradient: Ref.~\cite{BBDD2018pnasusa} finds that the mode of migration (persistent, bipedal, random) strongly depends on substrate stiffness. Corresponding reviews are \cite{GGDB2012s,MoYa2018np,AlTr2021pt,PaMi2023ebjbl}. Note that also the behavior of individual cells is sometimes discussed as a spreading phenomenon \cite{FRRA2010sm,PaMu2026arcmp} (Fig.~\ref{fig:sketch-active}~(d)).

In the examples above, wetting phenomena are studied for active matter characterized by self-propelling constituents and/or active stresses that do not proliferate on the considered timescale. However, proliferation has to be taken into account when investigating the growth of bacterial colonies and biofilms on solid (or liquid) substrates. There, the production of wetting agents (biosurfactants) by bacteria is intensively discussed for some time already \cite{MaMa1993crm,BeCL2000ap,VBDF2008tm,ARKW2009pnasusa,KAHH2020pre}, see e.g., the resulting front instabilities of the biofilm spreading front in \cite{FPBV2012sm,TrJT2018sm}. In contrast, the role of wetting and adhesion properties, e.g., in osmotic biofilm spreading \cite{SAWV2012pnasusa,SrKM2019e} (Fig.~\ref{fig:sketch-active}~(g)), became more recently a point of interest \cite{YNSW2017nc,TJLT2017prl}. The above discussed observations for cell aggregate spreading are paralleled by similar phenomena studied for biofilms. So does a transition from continuous to arrested biofilm spreading occur when decreasing the biofilm wettability on a rigid substrate \cite{TJLT2017prl}. Furthermore, also for biofilms the substrate softness is an effective control parameter \cite{CRAP2020e,AHGC2022pn,PiJT2025sm,FaWP2025ab}. Note that also the onset of sliding motion of a drop of bacterial suspension on an incline is promoted by bacterial motility \cite{HTCD2017pnasusab}, i.e., active depinning may be considered in analogy to its passive counterpart \cite{BKHT2011pre}.

More recently, liquid-like biomolecular condensates (granules) that play important roles inside cells have become a focus of attention \cite{BeBH2018rpp, ZKQP2024cr,CUNF2025nrmcb}. In cases where the underlying phase separation processes depend on out-of-equilibrium exchange processes and chemical reactions, their adhesion to each other and to various membranes is discussed as an example of active wetting \cite{BECR2009s,BeBH2018rpp,MCZ2023nc}: Condensates exhibit a well-defined contact angle when sitting at the cell nucleus and may undergo different wetting transitions at membranes \cite{MMDC2025cc}. This is closely related to the chemically active wetting phenomena modeled in \cite{LZWJ2025pnasusa}. There, coupled decomposition and transport processes in sessile condensate drops and membrane substrate are considered for an active system marked by adsorption and desorption processes that break the detailed balance of the reaction rates.  Note that recently also the migration of such condensates has been studied \cite{GDMF2024prr} (though yet without a membrane substrate), in principle, relating these systems to cases of migrating cell aggregates. Note, however, that even without out-of-equilibrium chemical reactions a rich spectrum of (equilibrium and relaxational) wetting phenomena occurs \cite{KuMK2021jcb,WLMM2024n,MaDi2024arb,MMDC2025cc,LLVH2025jacs,FMSH2025copb,MSSH2025nc}.

Beside the mentioned studies of active wetting in specific biophysical systems another body of work focuses on wetting phenomena in minimal and generic theoretical models of active matter represented by self-propelled particles or by entities that consume energy to generate motion via active stress. For instance, Ref.~\cite{NTTD2021sm} uses a lattice model for self-propelled particles near a solid substrate (studied by kinetic Monte-Carlo simulation) to show that active particles can produce a wetting layer of diverging thickness even when particle-particle and particle-substrate interactions are purely short-range repulsive, i.e., only due to excluded volume. The observed active wetting transition is directly related to the occurrence of motility-induced phase separation (MIPS) in the bulk \cite{CaTa2015arcmp,Spec2016epjt,BeRZ2018pre,VrBW2023jpcm}, i.e., the wetting layer thickness diverges at the MIPS threshold.

Similar studies for active Brownian particles (ABP) at different substrates, e.g., a thin membrane \cite{TuWi2021prlb} and a repulsive barrier in wedge shape \cite{TuJW2024sm}, also show wetting transitions that on a phenomenological level closely resemble corresponding equilibrium continuous and discontinuous phase transitions, although the mechanisms are very different. Similar results are obtained in Refs.~\cite{SeSo2017prl,SeSo2018pre,PeSo2025pre} for a number of on- and off-lattice run-and-tumble particle models, e.g., analyzing transitions from complete to partial wetting with increasing tumble rate, also see corresponding results obtained with a dynamical density functional theory (DDFT) based on effective activity-induced particle-particle and particle-substrate interactions \cite{WiBr2016el}. Beyond the mentioned similarity to equilibrium wetting, there are aspects intrinsic to self-propelled particles, namely, that the wetting layer also show polarization fields (i.e., local averaged direction of self-propulsion) \cite{TuJW2024sm,PeSo2025pre} (Fig.~\ref{fig:sketch-active}~(h)) with a somewhat similar role as director orientations in capillary and wetting phenomena of liquid crystals. One might even argue that the transition from an ensemble of bacteria swarming at a solid substrate to a sedentary biofilm \cite{Anir2022nrp} is closely related to such polarization effects: Ref.~\cite{GrPA2021e} indicates than an occurring transition from a monolayer film to multilayer islands is initiated by jamming events  (local MIPS nuclei) and overall seems to resemble an active dewetting process.

There are also extensions to particles with enhanced rotational diffusion at the substrate \cite{DaCh2025jcp} and to particle mixtures \cite{RoDS2023epje,RoCS2023pre}. Note that the addition of inertia may reduce or eliminate the motility-induced wetting of substrates \cite{CBLS2024cp}. Recently, Ref.~\cite{GrJC2025arxiv} derived a hydrodynamic theory for the wetting by ABP in a slit geometry and showed that on the continuum level the (nonequilibrium) phase diagram shows macroscopic complete and partial wetting states that on a phenomenological level behave quite similar to the corresponding passive case, again with the additional polarization field. Note, however, that activity implies (as also described in many of the other works) that the emerging steady states still feature stationary density currents. Another related work \cite{ZZDK2026np} on ABP derives an active equivalent of Young's law. It takes the emergence of stationary currents into account, and furthermore shows that partially wetting drops of ABP are not scale-invariant and split above a threshold size, also see Ref.~\cite{SoZh2025cpl}. 

A comparison of the capillary rise of purely repulsive active particles against gravity in a tube as described by Monte Carlo simulations of a lattice gas model and time simulations of corresponding hydrodynamic equations coupling velocity and polarization dynamics shows identical behavior, again resembling corresponding equilibrium behavior \cite{WyRi2020prl}. Similar results are also obtained for active Brownian particles with repulsive particle-wall interactions, also in comparison to stationary states obtained with a Fokker-Planck description \cite{MCWR2024pre} (with a focus on the analysis of the stationary particle currents sustaining the steady meniscus). Such activity-induced capillary rise and the emergence of a Landau-Levich-type climbing film after an activity-induced transition from partial to complete wetting is observed in experiments with an active liquid containing microtubule filaments and kinesin molecular motors and as well in accompanying hydrodynamic simulations \cite{AKYW2022s}, see Fig.~\ref{fig:sketch-active}~(i).

Note that most studies mentioned up to here mainly focus on static configurations and their transitions when control parameters are changed. This directly represents an active equivalent of the equilibrium wetting of section~\ref{sec:equi} although many of the introduced models are fully dynamic and may, therefore, also be employed to study the relaxational dynamics towards the stationary nonequilibrium states. For instance, an early work studying a wetting-related capillary phenomenon involving active matter is Ref.~\cite{JoRa2012jfm}. It uses a simplified model of active nematohydrodynamics \cite{KJJP2004prl} to study the continuous activity-driven spreading of a sessile drop on a solid substrate. Such a process is an active-wetting equivalent of relaxational wetting discussed in section~\ref{sec:relax}. There exist several other studies of sessile drops of active matter that employ hydrodynamic models, e.g., for active nematics in a number of particular model variants, approximations, and geometries \cite{OKMB2005mmb,TjMC2012pnasusa,TTMC2015nc,KhAl2015pre,WhHa2016njp,LoEL2019prl,LoEL2020sm,TSJT2020pre,StJT2022sm,CoFT2023prr,LiQi2025pf,ChDa2025jfm}. Together, the various studies (that would merit a proper review) describe a rich behavior of sessile drops of active liquids - resting, spreading, migrating (Fig.~\ref{fig:sketch-active}~(e)), coalescing, splitting  (Fig.~\ref{fig:sketch-active}~(f)) drops driven by active liquids that self-propel and/or show active stresses. Sometimes, the drops are additionally driven by externally imposed gradients of wettability or activity -- resulting in active equivalents of settings of driven wetting as discussed in section~\ref{sec:driven}. However, considerations of the detailed relation between activity-influenced (effective) interface tensions and (static and dynamic) wetting behavior are still scarce and a systematic comparative consideration across different systems and modeling approaches that would allow one to extract general laws of dynamic active wetting remains a challenging task for the future.

The question of wetting properties, i.e., static and dynamic contact angles, of active media directly connects to the intensively discussed issue of (effective) interface energies and tensions in other active soft matter systems \cite{FTCN2021prl,LaOm2024pre,LaOm2025jcp,WeRF2026np}, for instance, active particles undergoing motility-induced phase separation, and mixtures with nonreciprocal interactions between the components. There, demixing processes, phase coexistence, interface characterization, routes from stationary to motile and oscillating states, the existence of bubbly phases and interfacial waves are all subject of much present discussion, see e.g.\ Refs.~\cite{CaTa2015arcmp,TjNC2018prx,SaAG2020prx,YoBM2020pnasusa,FrWT2021pre,DOCZ2023nc,BrMa2024prx,GCKY2024sm,ChEO2025pre,EvOm2025prr,DAGM2025prr,GLFT2025prl}. However, wetting phenomena, e.g., at solid substrates, are not yet systematically studied for such systems. In particular, when three or more phases can coexist, wetting phenomena become relevant even without a solid substrate. For instance, a passive ternary mixture with purely diffusive dynamics can show three-phase coexistence with interfaces that meet at angles given by Neumann's law. If such a system is made active by the introduction of nonreciprocity,\footnote{Although here we refer to nonreciprocal interactions between the components, i.e., to breaking Newton's third law, in the context of gradient dynamics descriptions of soft matter systems one may distinguish three main types of nonreciprocity. These are (i) mechanical nonreciprocity (breaking Newton), (ii) thermodynamic nonreciprocity (breaking Onsager’s reciprocity relations), and (iii) chemical nonreciprocity (breaking detailed balance in chemical reactions, here, briefly called \enquote{breaking de Donder}). Historically, the thermodynamically consistent treatment of chemical reactions is due to Marcelin \cite{Marcelin1914}.}
the Neumann angles may change and the three-phase contact region can show persistent translational or rotational motion \cite{MaCa2025njp}. Note that a nonequilibrium Neumann law and a wetting transition are also discussed for reactive multi-component protein systems described by mass-conserving reaction-diffusion models \cite{WeRF2026np}. Similarly, for active media that show gas, liquid and solid phases, in the vicinity of a triple point one can discuss three-phase contact and the properties of wetting layers, see, e.g., the descriptions via active Brownian particle dynamics \cite{OKGG2021prl} and via a higher-order active phase-field-crystal (PFC) model \cite{HSVT2025prl}. However, the inclusion of such cases results in a rather strong widening of the spectrum of active wetting as it then encompasses the burgeoning field of mixtures with nonreciprocal interactions, certain reaction-diffusion models, etc.

\section{Conclusion and outlook}
\label{sec:conc}
%
The stated aim of this contribution has been to provide a tentative definition of \textit{active wetting} - a notion that has recently been employed to characterize a wide variety of phenomena. To lay a foundation for such a definition we have first proposed a classification of different cases of static and dynamic wetting into the categories of equilibrium wetting, relaxational wetting, driven wetting, and reactive wetting. Although at very first sight such a classification might look convincing, further scrutiny shows that one has to allow for caveats: In some cases, the distinction becomes ambiguous as it lies in the eyes of the beholder. For instance, a classification of a particular phenomenon as driven or reactive wetting (sections~\ref{sec:driven} and~\ref{sec:reactive}) instead of as relaxational wetting (section~\ref{sec:relax}) subtly depends on the employed conceptual idealizations and corresponding mathematical approximations, e.g., related to the relevant length and time scales, to the assumed infinite or finite spatial extension and boundary conditions of the considered domain, and to properties of the assumed reservoirs of energy or material. One may even say that without idealizations any dynamics would be relaxational.

However, even assuming that such (often implicit) idealizations and approximations can be taken for granted, major difficulties arise when considering static and dynamic wetting phenomena in systems that are permanently out of equilibrium: beyond the systems with imposed global spatial gradients along the substrate (broken parity), here termed \textit{driven wetting}  (section~\ref{sec:driven}), there exist many other systems involving sessile drops or three-phase contact lines (without imposed global gradients) that are permanently out of equilibrium, but not necessarily with moving contact lines or interfaces. Here we have termed all such systems reactive wetting as many of them are characterized by changes in state (thermodynamic phase or chemical state) and by implied couplings to infinite reservoirs (of energy or material). One way forward could be to use more descriptive categories, e.g., state transition-dominated wetting or transfer-dominated wetting, and then to split these into several sub-categories depending on the type of transition or through-flow. An objection is that this would create more confusion as it might escalate into subsequently finer and finer distinctions, e.g., one would need to separate wetting dominated by chemical reactions at the substrate, wetting dominated by chemical reactions at the liquid-gas interface, evaporation-induced wetting, etc. Instead, the here presented pragmatic approach uses a relatively straightforward coarse-grained classification and accepts that there is some overlap between categories (see examples discussed at the end of section~\ref{sec:driven}), and that there exist a few cases that fall through the gaps. For instance, we have not at all discussed any system with externally controlled spatio-temporal wettability patterns \cite{JoTh2007apl,GrSt2021sm,STGH2023prf}, e.g., switched by electrowetting \cite{MKBS2005jpcm}, by opto-electrowetting \cite{ZRDI2024ami}, or via substrates covered with photo-switchable layers \cite{HTRH2022jacs,NVGM2023aami,GAHS2023jpcc}. Closely related are also substrates with spatio-temporal topography patterns, e.g., substrates undulating due to elastic waves \cite{GrSt2024sm}, or uniform substrates that are vibrated \cite{DSGC2004l,BrED2007prl,NoKC2009prl,JoTh2010prl,DZGL2018jfm}. Another caveat is that even within the categories of driven and reactive wetting one can distinguish static and dynamic situations.

Finally, we come back to the case of active wetting (section~\ref{sec:active}) and propose a characterization that distinguishes the thereby addressed phenomena from the four earlier discussed categories. This is tricky as already the term \enquote{active} is used differently in different contexts: With \enquote{active system} one often refers to a rather wide spectrum of systems that are permanently out of equilibrium, i.e., normally (but not always), with a nonvariational mathematical description, while \enquote{active liquid} (or more general, \enquote{active matter} refers more specifically to a liquid (or some other kind of matter) that consists of self-propelled constituents and/or features active stresses resulting from properties of the constituents (see section~\ref{sec:intro}). Both draw on abundant homogeneously distributed energy sources (that are normally not explicitly considered) what corresponds to an idealized description. If, in contrast, the external energy source is made part of the consideration, e.g., by taking into account the spatio-temporal dynamics of chemical fuel and waste \cite{AAFE2024jcp}, and as well the specific mechanisms of chemo-mechanical coupling,\footnote{Here, chemo-mechanical coupling refers to the coupling between chemistry and mechanics, e.g., if chemical reactions of surfactants result in changes in surface tensions that cause Marangoni forces that drive mechanical (liquid) motion that vice versa influences the chemical reactions. For a recent example of dynamic wetting phenomena arising when the resulting feedback loop is driven out of equilibrium, see \cite{VoTh2025prf}.} one might see active wetting as a special case of reactive wetting. This is particularly appealing in cases where the chemical reactions are explicitly considered, e.g., when studying the wetting and adhesion behavior of biomolecular condensates at membrane substrates. In Ref.~\cite{LZWJ2025pnasusa} the chemical reactions that are permanently kept away from equilibrium correspond to binding processes at the substrate, and the occurring drop dynamics is termed chemically active wetting.

Going through the list of all examples mentioned in section~\ref{sec:active}, we think that it could be a clarifying restriction to use the term \textit{active wetting} only for wetting phenomena involving active liquids, i.e., where the chemo-mechanical coupling takes place on the level of the microscopic bulk constituents. This then includes most examples given in section~\ref{sec:active}, namely, the (de)wetting phenomena for cell monolayers and aggregates, for dense clusters and layers of active Brownian particles, and for sessile drops of active liquids. However, the category of active wetting would then neither include biomolecular condensates if only diffusive transport is accounted for nor proliferating matter like biofilms - these would more naturally fit into the category of reactive wetting where then also wetting phenomena involving nonreciprocal media would belong, e.g., in case one investigated them with nonreciprocal Cahn-Hilliard models that are purely diffusive. However, similar caveats as discussed above also apply here, e.g., regarding the distinction of reactive and active wetting: Again, there are systems where the classification depends on the level of description. For instance, a system could show active wetting when described microscopically via active Brownian particles and run-and-tumble particles and reactive wetting when described macroscopically via a coarse-grained nonreciprocal field theory \cite{DiOT2024jpamt}.\footnote{Note that \cite{DiOT2024jpamt} does not discuss wetting but the different levels of description.} A wetting-related example where something similar occurs with respect to the classification as relaxational are sessile drops on a heated substrate - a system that is clearly permanently out of equilibrium. However, in long-wave approximation \cite{OrDB1997rmp} its evolution toward a steady state is described by a thin-film equation in gradient dynamics form where the ``energy functional'' depends on heat flux and resulting Marangoni force, see \cite{OrRo1992jpif} and section~3 of \cite{ThKn2004pd}. In this case, the approximate description corresponds to one of relaxational wetting.

  Furthermore, the listing in section~\ref{sec:active} indicate that also for active wetting there exist variants that mirror many of the phenomena and set-ups of the earlier discussed cases: Typical subjects are, for instance, the dependence of the thickness of uniform wetting layers on control parameters (with the additional aspect of polarization), the shape and contact angle of static drops (now representing force balances instead of thermodynamic equilibria), the advance of films in a Landau-Levich setting, and the speed, dynamic contact angle and stability of moving drops. This is reflected, e.g., in the notion of \enquote{active equilibrium wetting}  used in Ref.~\cite{ChDa2025jfm}. Although, to us this particular notion seems contradictory and therefore problematic, we agree that an important focus should be on the identification of similarities and differences between phenomena of static and dynamic wetting in the various described settings and phenomenologically similar cases of on the one hand equilibrium and relaxational wetting, and on the other hand driven, reactive and active wetting. In this way, one might be able to establish specific mappings between the laws capturing dependencies of, e.g., static and dynamic contact angles on the various relevant control parameters across several categories.

To summarize, we have first introduced a tentative classification of wetting phenomena into the categories of equilibrium, relaxational, driven, reactive and active wetting, but have then discussed a number of caveats, exceptions and ambiguities that are deeply related to the type and level of employed modeling idealizations. On the one hand, this indicates that we should always clearly state what we refer to when discussing \textit{active wetting}. On the other hand, we believe that future systematic comparative studies across different systems will allow for an assessment and improvement of taxonomies like the present tentative proposal. On an epistemological level \cite{KnuuttilaCarrilloKoskinen2025}, we optimistically think that the described caveat-riddled and shape-shifting classification should not be a source of despair as exactly such transitions in categorization resulting when passing though a hierarchy of modeling approaches may themselves be a source of more complete classifications of the underlying physical systems.

\acknowledgments

The work has benefited from discussions with colleagues at M\"unster and beyond, namely, members of the group \textit{Self-organization and Complexity}, members of the Priority Program SPP~2171 \textit{Dynamic Wetting of Flexible, Adaptive, and Switchable Surfaces}, and many participants of the below mentioned programs at INI, KITP and MPIPKS. Further, I thank the Deutsche Forschungsgemeinschaft (DFG) for support via Grants TH781/12-1, and TH781/12-2 within SPP~2171; the Isaac Newton Institute (INI) for Mathematical Sciences, Cambridge, for support and hospitality during the program \textit{Anti-Diffusion in Multiphase and Active Flows (ADIW04)} (2025, supported by EPSRC grant EP/R014604/1 to INI); the Kavli Institute for Theoretical Physics (KITP), Santa Barbara, for support and hospitality during the program \textit{Active Solids} (2025, supported in part by grant NSF PHY-2309135 to KITP);  the Max Planck Institute for the Physics of Complex Systems (MPIPKS) for support and hospitality during the workshop \textit{From Adaptive to Active Wetting Dynamics} (2026); and the doctoral school \enquote{Active living fluids} funded by the German French University (Grant CDFA-01-14).

\end{document}